\documentclass[10pt,twocolumn]{article}
\pdfoutput=1
\usepackage{amsmath}
\usepackage{amssymb}
\usepackage{url}
\usepackage{graphicx}
\usepackage{cite}
\usepackage{chicago}
\usepackage{hyperref}
\hypersetup{
pdfauthor=Havemann;  Gl\"aser; Heinz; Struck,
pdftitle=Havemann et al. (2012): Evaluating Overlapping Communities
with the Conductance of their Boundary Nodes,
colorlinks=true,
linkcolor=blue,
anchorcolor=blue,
citecolor=blue,
urlcolor=blue,
menucolor=black
}

\usepackage{color} 

\title{Evaluating Overlapping Communities\\
with the Conductance of their Boundary Nodes}
 \date{}

\author{Frank Havemann$^{1,\ast}$  \and Jochen Gl{\"a}ser$^{2}$  \and Michael Heinz$^{1}$ \and Alexander Struck$^{1}$}
\begin{document}
\maketitle
\begin{flushleft}

{1} Institut f{\"u}r Bibliotheks- und Informationswissenschaft, Humboldt-Universit{\"a}t zu Berlin, Berlin, Germany 
\\ 
{2} Zentrum Technik und Gesellschaft, Technische Universit{\"a}t Berlin, Berlin, Germany
\\ 
$\ast$ E-mail: Frank (dot) Havemann (at) ibi.hu-berlin.de
\end{flushleft}

\section*{Abstract}

Usually the boundary of a community in a network is drawn between
nodes and thus crosses its outgoing links. If we construct
overlapping communities by applying the link-clustering approach 
nodes and links interchange their roles. Therefore, boundaries must drawn through the nodes shared by two or more communities. 
For the purpose of community evaluation we define a 
conductance of boundary nodes of overlapping communities
analogously to the graph conductance of boundary-crossing links used
to partition a graph into disjoint communities. 
We show that conductance of boundary nodes (or \textit{normalised node cut}) can be deduced from
ordinary graph conductance of disjoint clusters in
the network's weighted line graph 
introduced by \citeN{evans2009line} to get overlapping
communities of nodes in the original network. 
We test whether our definition can be used 
to construct meaningful overlapping communities with a local
greedy algorithm of link clustering. In this note we 
present encouraging results we obtained for Zachary's
karate-club network. 

\section{Introduction}

A community in a network is usually defined as a subgraph 
that is both cohesive and well separated from the rest of the network \cite{Fortunato2010community}. Cohesion and separation can be evaluated by various measures. A simple absolute measure of cohesion is the sum of weights of links between all members $C$ of a subgraph  which equals their total internal degree $k_\mathrm{in}(C)$ divided by 2. A simple absolute measure of separation is the sum of weights of links between members and non-members which equals the total external degree $k_\mathrm{out}(C)$.  A single function sensitive to a subgraph's cohesion and separation is the \textit{normalised cut}
$$\Phi(C) = \frac{k_\mathrm{out}(C)}{k_\mathrm{in}(C)+k_\mathrm{out}(C)}.$$
The total external degree $k_\mathrm{out}(C)$ equals the total electrical conductance of the links cut by $C$'s boundary if each link's conductance is defined by its weight. 
Therefore $\Phi(C)$ is also named \textit{conductance}.\footnote{$\Phi(C)$ is called 
conductance only if the total degree of $k(C)=k_\mathrm{in}(C)+k_\mathrm{out}(C)$ is smaller than the total degree of $C$'s complement, s.\ the review by \citeN{Fortunato2010community} and references therein.} 
If the cut through external links of a subgraph with node set $C$ has minimal conductance $\Phi(C)$ then $C$ can be called a community. \citeN{yang2012defining} tested 13  evaluation functions and found that evaluating subgraphs with conductance $\Phi$ results in good disjoint communities.

These ideas can also be applied when the perspective on the network is changed and communities of links instead of communities of nodes are to be constructed. This approach was introduced by \citeN{evans2009line} and by \citeN{ahn2010link} with the aim to obtain overlapping node communities.\footnote{Link clustering is advantageous for some tasks of community detection, for example, if communities representing thematic structures in networks  of papers are to be constructed, the focus on links is to be preferred because citation links between papers are thematically more homogenous than papers themselves \cite{havemann_identifying_2012}.} The definition of a community as a cohesive and separated subgraph still holds for this perspective. The important difference between communities of nodes and communities of links is that the latter's boundaries cuts through nodes instead of links. 
When community boundaries shift from links to nodes, the evaluation function needs to be changed accordingly. In this paper, 
we introduce the \textit{normalised node cut} $\Psi$ as an evaluation function for link communities. 

\citeN{evans2009line} found that a network's line graph can be used to obtain link communities. They applied modularity---a global evaluation function---to obtain link communities but stressed that any method for community construction can be applied to the line graph.
We show that evaluating a network's subgraphs  with the normalised node cut $\Psi$  
is equivalent to an evaluation of subgraphs in the network's line graph
with ordinary normalised edge cut $\Phi$ if each edge in the line graph is weighted with the degree of the corresponding node as  proposed by \citeN{evans2009line}.

We test whether the normalised node cut $\Psi$ can be used to evaluate subgraphs and thus to find a network's link communities. For this purpose we construct a $\Psi$-landscape. A community of links is defined as a subgraph with a local $\Psi$-minimum.
We apply a greedy local expansion algorithm to find local minima in the $\Psi$-landscape. We present results obtained for a simple benchmark, the karate-club network analysed by \citeN{zachary1977information}. 

\section{Method}
\subsection{Normalised Node Cut}
Since the boundary of a link community consists of nodes, the measure of cohesion and separation must be shifted accordingly. We define the \textit{normalised node cut}  $\Psi(C)$ of a connected subgraph with node set $C$ as the normalised total conductance of $C$'s boundary nodes given by
\begin{equation}\begin{split}
\Psi(C) & = \frac{1}{k_\mathrm{in}(C)}\sum_{i \in C}\frac{1}{1/k_i^\mathrm{in}(C) + 1/k_i^\mathrm{out}(C)}\\ & = \frac{1}{k_\mathrm{in}(C)}\sum_{i \in C}\frac{k_i^\mathrm{in}(C) k_i^\mathrm{out}(C)}{k_i}.
\end{split}
\label{eq:Psi}
\end{equation} 
Here $k_i^\mathrm{in}(C)$ is the internal degree of node $i$, i.e.\ the total weight of its links to other $C$-members, and $k_i^\mathrm{out}(C)$ its external degree, i.e.\ the total weight of its links to non-members. Both sum up to the total degree of node $i$, which does not depend on $C$:
$k_i = k_i^\mathrm{in}(C)+ k_i^\mathrm{out}(C)$. 

Each term in the sum of equation \ref{eq:Psi} equals the electrical conductance between external and internal nodes connected through node $i$ if we identify the link weights with electrical conductances. Since the external degree $k_i^\mathrm{out}(C)$ of inner nodes is zero, the sum includes only the nodes that constitute $C$'s boundary. 

The normalised node cut $\Psi$ is defined for weighted networks. 
For the sake of simplicity, we restrict the discussion to unweighted networks in the remainder of the paper.
In this case, $k_i^\mathrm{in}(C)$ equals the number of links between node $i$ and other members of $C$ and  $k_i^\mathrm{out}(C)$ is the number of links between $i$ and nodes outside $C$.

For the complete graph, which has no outgoing links, $\Psi=0$ because for all nodes $k_i^\mathrm{out}=0$. 
Furthermore, the normalisation in equation \ref{eq:Psi} guarantees that $\Psi < 1$ for all subgraphs because $\Psi$ equals the $k_i^\mathrm{in}$-weighted average of relative external degrees $k_i^\mathrm{out}/k_i < 1$.\footnote{One of the 13 evaluation functions tested by \citeN{yang2012defining} is the \textit{unweighted} average of relative external degrees (they call it \textit{average out degree fraction}).} 

Function $\Psi(C)$ decreases with increasing internal cohesion (measured by $k_\mathrm{in}(C)$) and with decreasing linkage with the rest of the network (measured with the sum in equation \ref{eq:Psi}). Thus, $\Psi(C)$ 
is a function  sensitive to a subgraph's cohesion and separation.
 
Our definition of the normalised node cut $\Psi$ can be derived by applying the normalised edge cut $\Phi$
in the network's line graph with weights $1/k_i$. This weighting was proposed by \citeN{evans2009line}. To construct a network's line graph we first define an auxiliary bipartite graph obtained by putting a node on each link of the original network. The affiliation matrix $B$ of the bipartite graph---also called its incidence matrix---has a row for each of the $n$ original nodes and a column for each of the $m$ original links. Each link column contains only two non-zero elements, namely the elements in the rows of the nodes $i$ and $j$ connected by the link. We can project the bipartite graph back onto the original network with the product $BB^\mathrm{T}$ which equals its adjacency matrix $A$  (except for the main diagonal).

We obtain the network's line graph by the opposite projection $B^\mathrm{T}B$ of the bipartite graph. \citeN{evans2009line} emphasise, that the line graph contains the same amount of information as the original network in all cases of practical interest. Knowing $B^\mathrm{T}B$ we can almost always calculate $BB^\mathrm{T}$ and thus also the network's adjacency matrix $A$.

\citeN{evans2009line} weight the edges of the line graph with the inverse degree $1/k_i$ of the node $i$ in the original network because each node is represented as a clique in the line graph. They define the line graph's adjacency matrix as
\begin{equation}
E_{kl} = \sum_{i=1}^n \frac{B_{ik}B_{il}}{k_i}.                                               
\label{eq:E}
\end{equation} 
For this line graph we can calculate the ordinary graph conductance or normalised cut $\Phi$ of a link set $L$ and get $\Phi(L) = \Psi(C(L))$, where $C(L)$ is the set of nodes attached to links in $L$. The proof can be found in the appendix (p. \pageref{appendix}).

Different link sets $L$ can have the same induced node set $C(L)$ if we define $C(L)$ to be the set of all nodes attached to links in $L$. We define the link set $L(C)$ induced by $C$ as the maximum  set of links (existing in the network) that induces $C$. A connected subgraph's link set is assumed to be a maximum set. It is induced by the subgraph's node set. If we would not include all existing links between all nodes of a subgraph in its link set we would have external links between member nodes. We could even change two adjacent inner nodes to boundary nodes if we would omit the link connecting them from the subgraph's link set. 

Weighting the line graph's edges with the inverse degrees of  nodes in the original network is equivalent to an Euclidean normalisation of the nodes' vectors in the affiliation matrix $B$ of the auxiliary bipartite graph. This becomes clear if we factorise the terms of the sum  in equation \ref{eq:E}:  
\begin{equation}
E_{kl} = \sum_{i=1}^n \frac{B_{ik}}{\sqrt{k_i}}\frac{B_{il}}{\sqrt{k_i}}.
\label{eq:factorise}
\end{equation}
Then we can shortly write $E=D^\mathrm{T}D$ with $D_{ik}=B_{ik}/\sqrt{k_i}$ and verify the Euclidean normalisation of the $n$ row vectors of $D$:
\begin{equation}
\sum_{k=1}^m D_{ik}^2=  \sum_{k=1}^m\frac{B_{ik}^2}{k_i} =  \frac{1}{k_i}\sum_{k=1}^m B_{ik}=1.
\label{eq:L2}
\end{equation}
Here we used that $B$ is binary (because $A$ is binary for unweighted networks) and therefore $B_{ik}^2=B_{ik}$.

The projection of the normalised bipartite graph described by affiliation matrix $D$ back on a network of the original nodes is given by $DD^\mathrm{T}$. Any element of adjacency matrix $DD^\mathrm{T}$ (except for the main diagonal)
is given by 
\begin{equation}
\sum_{k=1}^m D_{ik}D_{jk} = \sum_{k=1}^m\frac{B_{ik}B_{jk}}{\sqrt{k_ik_j}} =  \frac{A_{ij}}{\sqrt{k_ik_j}}.
\label{eq:back-projection}
\end{equation}
This means that the Euclidean normalisation of $B$'s row vectors is equivalent to weighting each link in the original (unweighted) network with the geometric mean of its nodes' inverse degrees.  The weighted graph described by adjacency matrix $E$ is not the line graph of the unweighted network  described by adjacency matrix $A$ but the line graph of the network weighted according to equation \ref{eq:back-projection}. The approach is applicable only to those networks for which the weighting with the geometric means is a realistic assumption. 

\subsection{Defining Communities}

Local greedy algorithms construct communities by starting from seeds and adding those neighbours of the subgraph which maximally improve or minimally downgrade its cohesion and separation.
Due to its locality, this approach can be used to construct a seed's nested sequence of communities in a network which is too large to process it totally~\cite{clauset2005flc,luo_exploring_2008,havemann2011identification}. 
In the case of link communities, the local greedy expansion of subgraphs should start from links as seeds and iteratively add new links.

We construct a landscape of the normalised node cut $\Psi(C)$ of connected subgraphs.  Each connected subgraph consists of a node set $C$ and the link set $L(C)$ containing all links that exist between nodes in $C$. 
Each place in the landscape represents a connected subgraph defined by its node set $C$. A relation between two places exists if  one of the subgraphs can be obtained by adding a node to the other one. The height of a place is the subgraph's $\Psi$-value.\footnote{If we imagine the relations to be located on a two-dimensional surface we cannot avoid that they cross each other. A better imagination is therefore that relations are like cable-ways between places of different height.} 
Communities can be defined as those subgraphs whose $\Psi$-values are local minima in the $\Psi$-landscape.\footnote{This landscape concept is not restricted to link communities but can be used with all local evaluation functions for communities.}
 
Any two places in the $\Psi$-landscape are connected by paths. To reach one place from another one we have to add and remove nodes of the corresponding subgraphs. The absolute distance between two places  in the $\Psi$-landscape is the number of steps one has to go on a shortest path between them. These are $|M \cup N| - |M \cap N|$ steps, where  $M$ and $N$ are the node sets of the two connected subgraphs. We obtain the Jaccard distance by normalising the absolute distance with  $|M \cup N|$.

The distance between two communities can be used to define a community's stability. A community's stability is the shortest Jaccard distance to a community with a lower $\Psi$-value. A community is more cohesive and better separated from the rest of the network than all other communities within the radius of the shortest Jaccard distance to a community with a lower  $\Psi$-value. 

The communities evaluated by $\Psi$ are connected subgraphs that grow by adding neighbouring nodes with all their links to the community. However, it is justified to treat these communities as link communities because the community's boundary consists of nodes. Expanding a link community means shifting its boundary from one set of nodes to another, while the node communities commonly discussed in the literature are expanded by shifting their boundaries from one set of links to another.

\subsection{Identifying Communities \\with a Greedy Algorithm}

Constructing the whole $\Psi$-landscape is not feasible for larger networks. However, algorithms for the search of stable local minima in the $\Psi$-landscape can be constructed. We tested a greedy algorithm that adds those nodes that incur either the greatest reduction in $\Psi$ (i.e.\ go down the steepest slope in the $\Psi$-landscape) or the smallest increase in $\Psi$ (i.e.\ go up the gentlest slope in the $\Psi$-landscape).

Starting from a seed link, we go downhill in the $\Psi$-landscape on the path with the steepest slope. This slope is produced by adding to the subgraph the neighbouring node that incurs the greatest reduction in $\Psi$. If the $\Psi$-balance of two candidate nodes is tied we randomly select a path. Experiments showed that in most cases the two paths created by nodes with tied $\Psi$-balance will merge soon. If adding any neighbouring node to a subgraph increases $\Psi$, we search for members of the subgraph whose \textit{exclusion} further reduces $\Psi$. If we don't find such members then we have reached a local $\Psi$-minimum. If we do, we prune the subgraph by excluding them and try again to add a neighbour which maximally reduces $\Psi$. If this trial fails then we have also reached a minimum.

After a local minimum has been reached, we continue to add nodes. Initially, nodes producing the smallest increase in $\Psi$ must be added for the algorithm to leave the hollow in the $\Psi$-landscape created by the local minimum. Thereafter, the search for the steepest slope can be resumed. This is repeated until we reach the ground state of the whole~(connected) network with $\Psi=0$. If all links of a network are used as seeds, there is a high likelihood that all local minima are found or at least all those with high stability.


For larger networks it takes much time to determine the path for each link. 
It is possible to save computing time by an updating of node sets and variables needed for the iterative procedure. 
In appendix \ref{app.PsiMin} we describe how the sum in the $\Psi$-function can be updated during the iteration.

\section{Experiments}
\subsection{Karate Club}
The club of 34 karate fighters observed by \citeN{zachary1977information} split up into two disjoint parts of equal size. The links between the fighters were weighted with their interactions at different places. The network of the karate club became a benchmark graph that is often used for testing cluster algorithms, including algorithms for the construction of overlapping communities. In our experiments, we used the unweighted version of the network.

We applied the greedy algorithm described above to each of the 78 links of the network. The algorithm found seven local minima in the network's $\Psi$-landscape. Table \ref{tab:communities} lists for each minimum the number of links, the number of nodes, the normalised node cut $\Psi$, and the number of seed links from which the minimum has been found.

\begin{table}[!b]
\caption{Communities found in the karate network}
\begin{center}
 \begin{tabular}{rrrrr}
name & links & nodes & $\Psi$ & seeds   \\
\hline
$C_1$ & 68 & 29 & .022 & 68 \\
$C_2$ & 43 & 21 & .077 & 40 \\
$C_3$ & 41 & 19 & .091 & 10 \\
$C_4$ & 10 & 6  & .150 & 10 \\
$C_5$ & 6 & 5  & .294 & 7  \\
$C_6$ & 2 & 3  & .460 & 2 \\
$C_7$ & 1 & 2  & .469& 1\\
\hline
 \end{tabular}
\end{center}

 \label{tab:communities}
  \end{table}

\begin{figure}[!t] 
\begin{center} 
\includegraphics[width=3in]{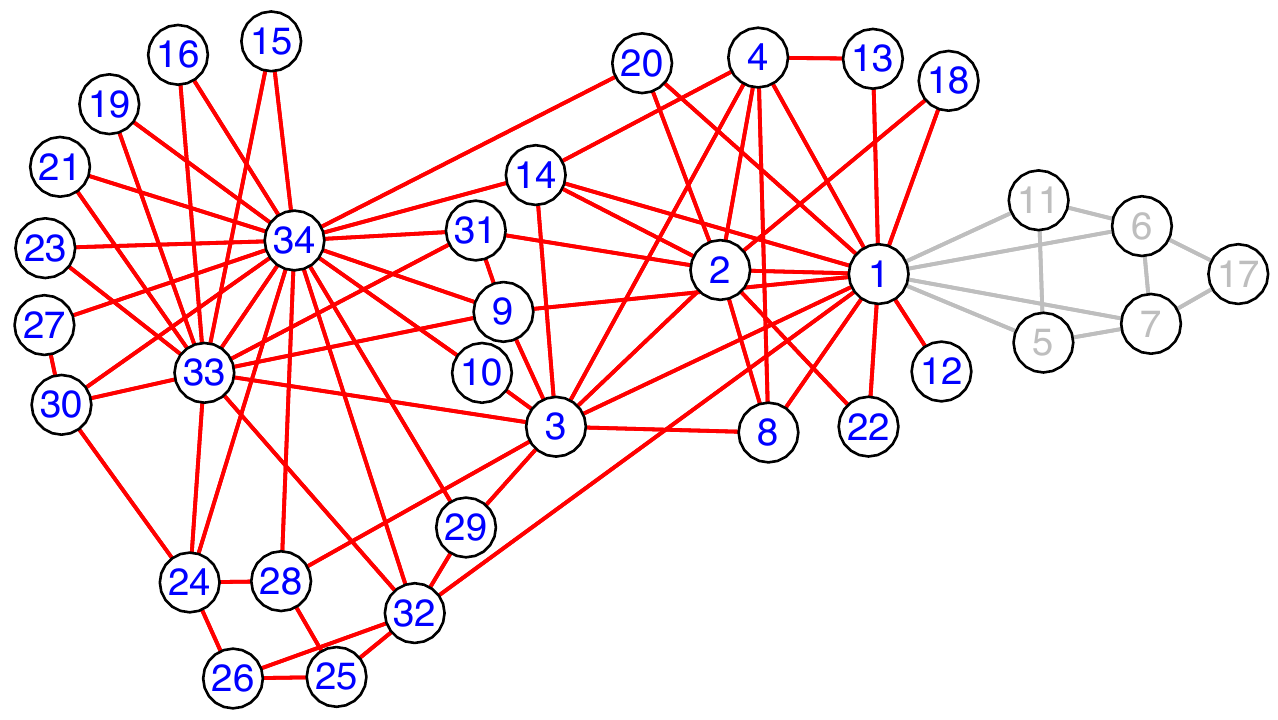} 
\end{center} 
\caption{Community $C_1$ has 29 nodes (blue) and 68 links (red). The ten grey links of community $C_4$ connect six nodes including the boundary node 1.} 
\label{Fig-karate-A} 
\end{figure}

\begin{figure}[t] 
\begin{center} 
\includegraphics[width=3in]{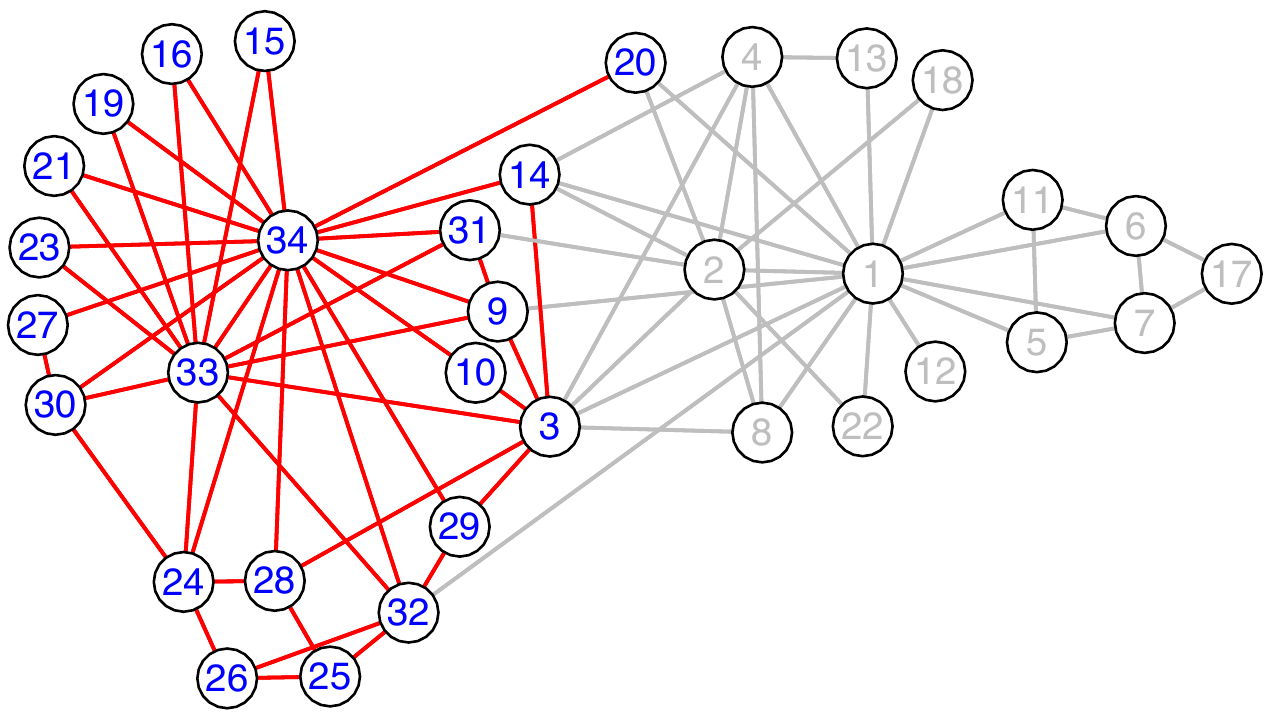} 
\end{center} 
\caption{Community $C_2$ has 21 nodes (blue) and 43 links (red). Community  $C_3$ has 19 nodes (connected by grey links). Both communities share six boundary nodes (connecting grey and red links).} 
\label{Fig-karate-B} 
\end{figure}

\begin{figure}[!t] 
\begin{center} 
\includegraphics[width=3in]{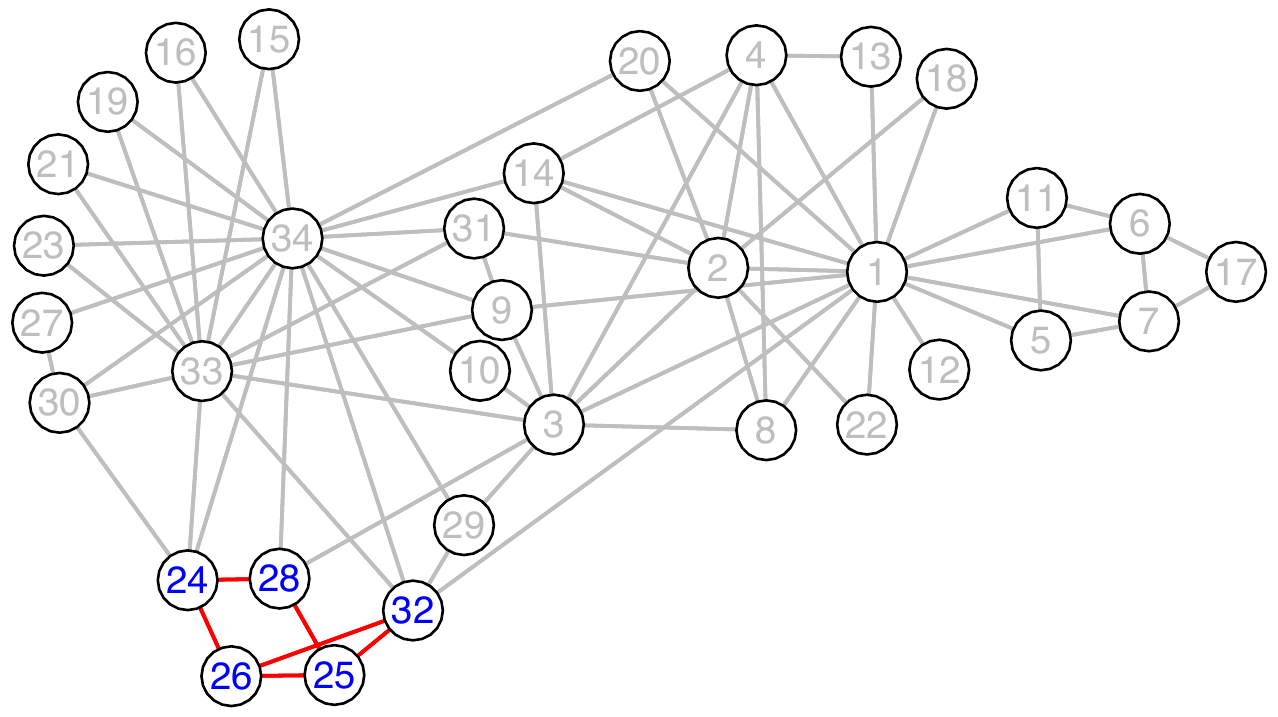} 
\end{center} 
\caption{Community $C_5$ has five nodes (blue) and six links (red).} 
\label{Fig-karate-E} 
\end{figure}

\begin{figure}[p] 
\begin{center} 
\includegraphics[width=3in]{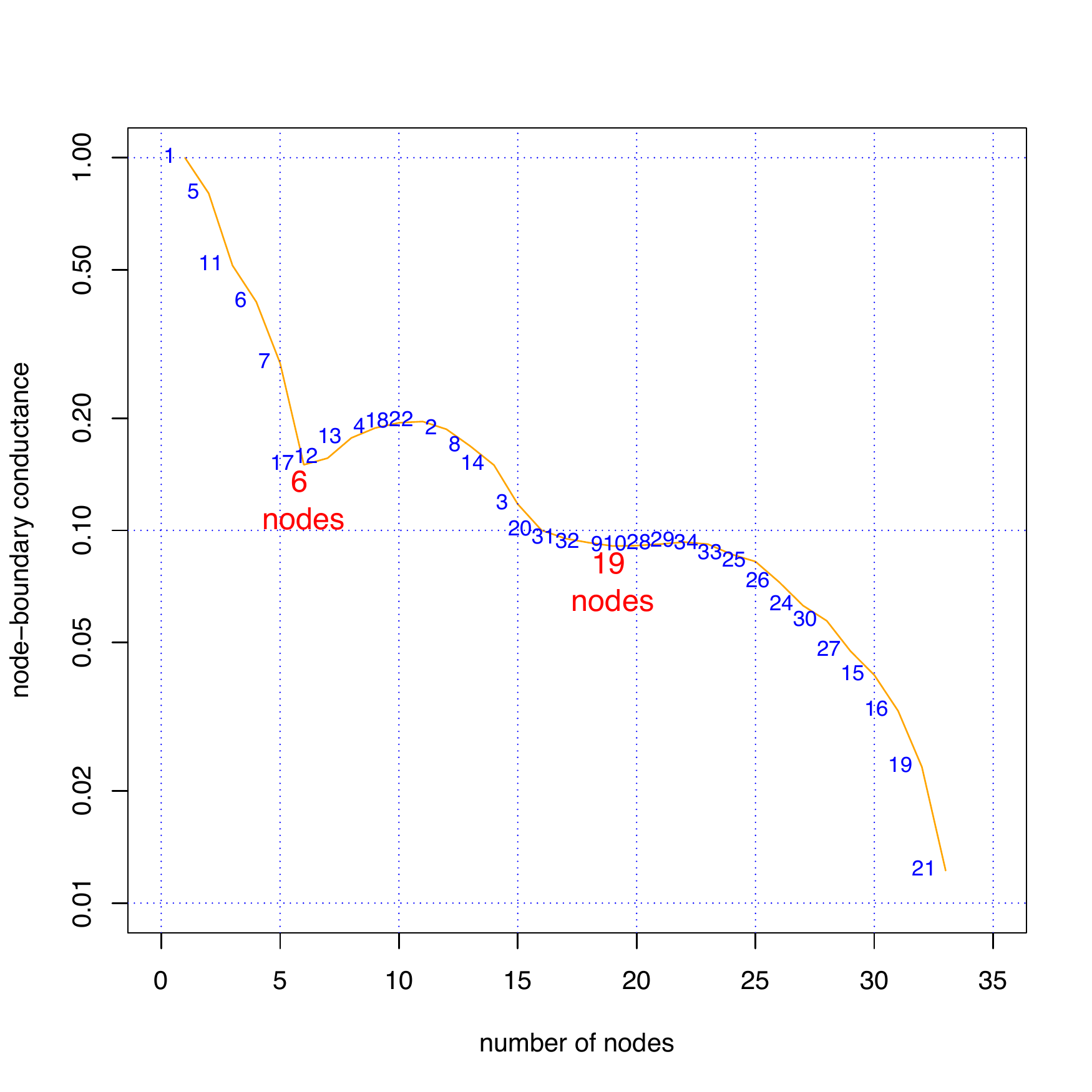} 
\end{center} 
\caption{$\Psi$-plot of seed link (1, 5) with minima of communities $C_4$ and $C_3$} 
\label{Fig-karate-1-5} 
\end{figure}

\begin{figure}[p] 
\begin{center} 
\includegraphics[width=3in]{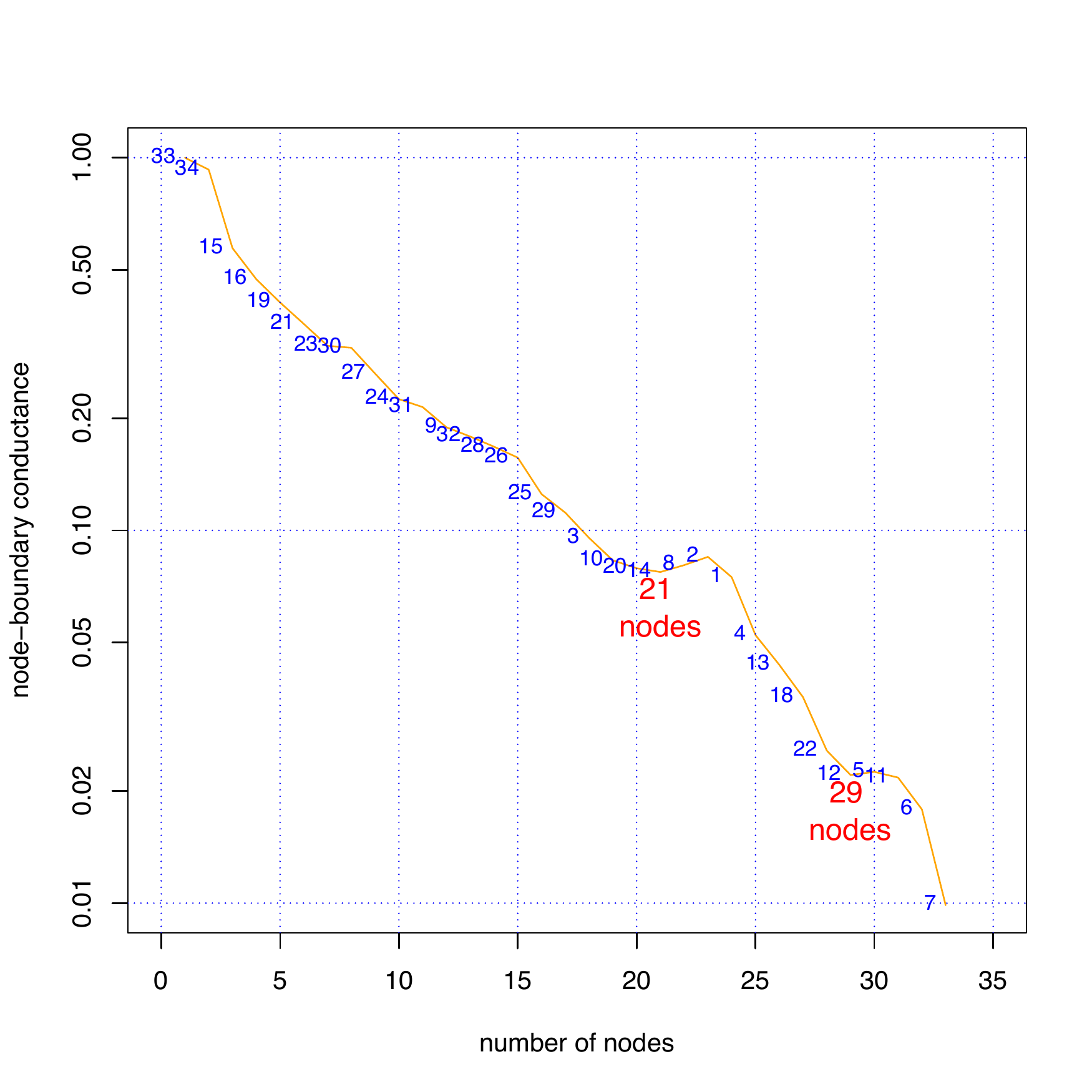} 
\end{center} 
\caption{$\Psi$-plot of seed link (33, 34) with minima of communities $C_2$ and $C_1$} 
\label{Fig-karate-33-34} 
\end{figure}

\begin{figure}[p] 
\begin{center} 
\includegraphics[width=3in]{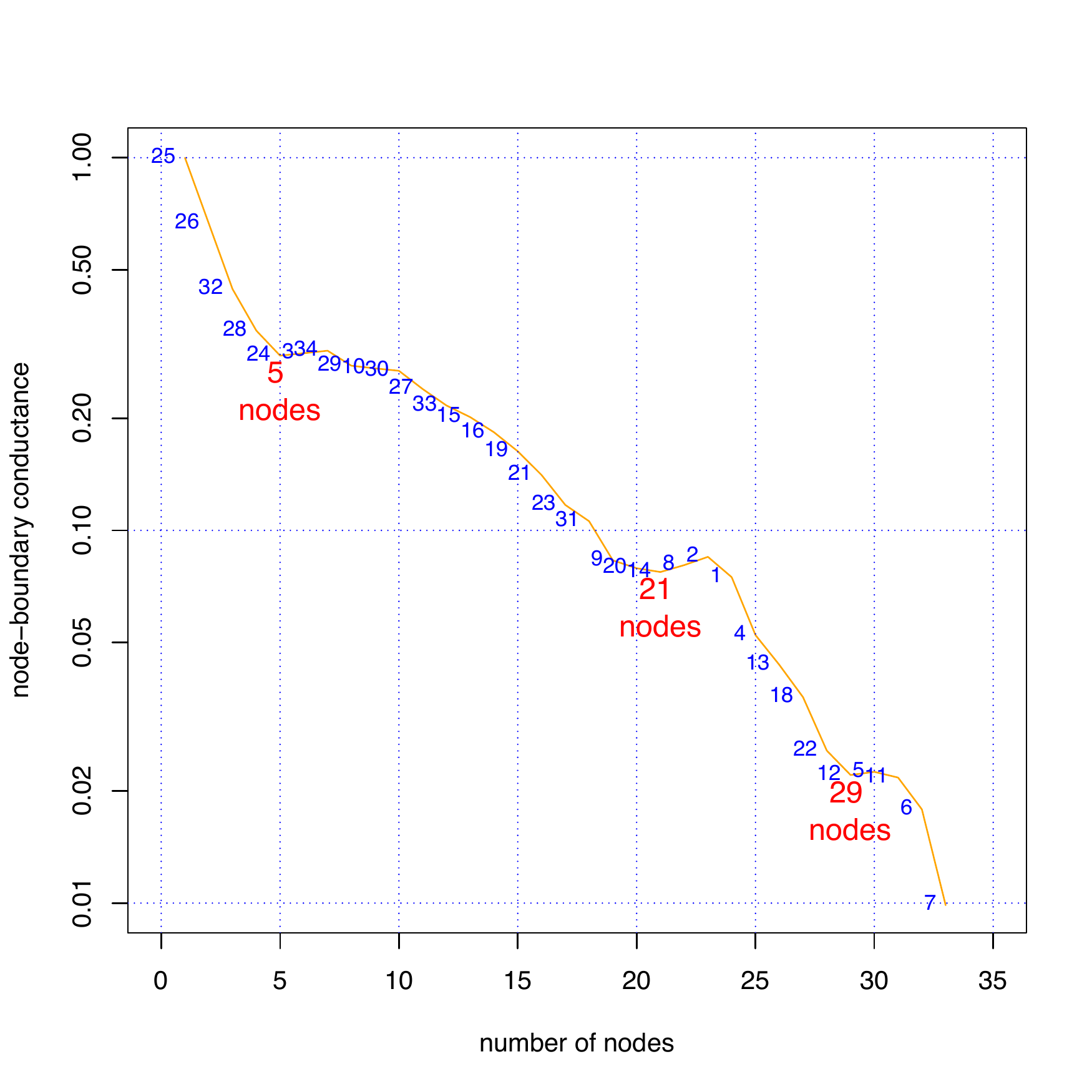} 
\end{center} 
\caption{$\Psi$-plot of seed link (25, 26) with minima of communities $C_5$, $C_2$, and $C_1$} 
\label{Fig-karate-25-26} 
\end{figure}

\begin{figure}[p] 
\begin{center} 
\includegraphics[width=3in]{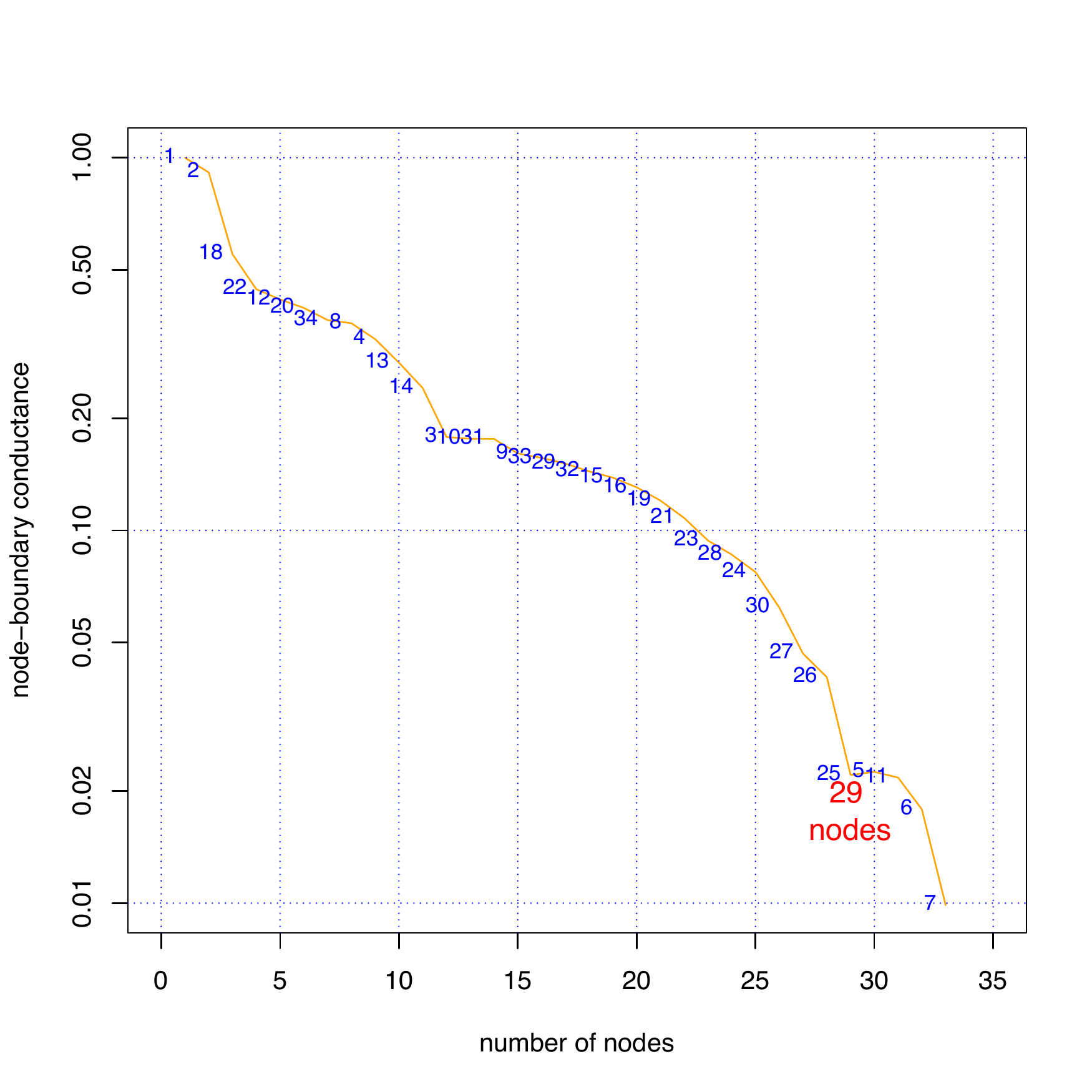} 
\end{center} 
\caption{$\Psi$-plot of seed link (1, 2) with minimum of community $C_1$} 
\label{Fig-karate-1-2} 
\end{figure}

The largest community $C_1$ with 29 nodes has only one boundary node, namely node 1 (s.\ Fig.~\ref{Fig-karate-A}). The same one-node boundary delineates community $C_4$ with six nodes. The union of $C_1$ and $C_4$ covers the whole network, their intersection contains only node~1. Analogously, the union of $C_2$ and $C_3$ also covers the network. Their intersection contains their common boundary nodes, namely nodes 3, 9, 14, 20, 31, and 32 (s.\ Fig.\ \ref{Fig-karate-B}). The other three communities are subsets of larger communities, $C_5$ and $C_6$ of $C_2$, and $C_7$ of $C_1$ and $C_3$. Community $C_5$ has five nodes as members (s.\ Fig.\ \ref{Fig-karate-E}), $C_6$ only three (nodes 3, 10, and 34). Community $C_7$ is the pair of nodes 1 and 12.

At least one local minimum of $\Psi$ was found in each run. 27 of the 78 seed links led to one local minimum, 42 to two local minima, and nine to three local minima. All ten links of community $C_4$ led to $C_3$ and $C_4$. As an example of the paths through the $\Psi$-landscape, Fig.\ \ref{Fig-karate-1-5} shows the plot of $\Psi$ over the number of nodes obtained by starting from seed link $(1,5)$. Since the $\Psi$-scale in Fig.\ \ref{Fig-karate-1-5} is logarithmic, the last point of the $\Psi$-curve, which represents the whole network with $\Psi=0$, is not visible.

40 of 43 links of $C_2$ found $C_1$ and $C_2$. As an example, we show the $\Psi$-plot of seed link (33, 34) in Fig.~\ref{Fig-karate-33-34}. The two links of $C_6 \subset C_2$ found also $C_6$, the six links of $C_5 \subset C_2$ also $C_5$. In Fig.~\ref{Fig-karate-25-26} we plot $\Psi$ of seed~(25,~26). 

Community $C_6$ is also present as a local minimum when we start from seed (3, 28). However, in this case node 3 is excluded in the pruning process because excluding it further reduces $\Psi$. Community $C_7$ is identical to its seed link (1, 12). All other 27 links led the algorithm to only one local minimum, namely that of $C_1$. All these links belong to $C_3$. As a further example, we plot the $\Psi$-curve of seed (1,  2) in Fig.~\ref{Fig-karate-1-2}.

Two pairs of communities found by the greedy algorithm each cover the whole network and overlap in their boundary nodes, namely the pair $C_2$ and $C_3$ and the pair $C_1$ and $C_4$. The cut by the boundary nodes between $C_2$ and $C_3$ is compatible with the final split of the karate club and corresponds to one solutions found by \citeN{evans2009line}.
Four of the six fighters on the boundary between $C_2$ and $C_3$~(nodes 3, 9, 14, 20) decided to follow Mr.\ Hi~(node~1) and the other two~(nodes 31 and 32) joined the officer's club~(node 34). 

Note that not only $C_2$ and $C_3$ overlap but also their link sets $L(C_2)$ and $L(C_3)$: $L(C_2) \cap L(C_3) = \{(3, 9), (9, 31)\}.$
This is a special feature of our greedy algorithm  based on the normalised node cut. The link clustering procedures proposed by \shortciteN{ahn2010link} and by \citeN{evans2009line} produce only disjoint link clusters.

\begin{figure}[!t] 
\begin{center} 
\includegraphics[width=1.7in]{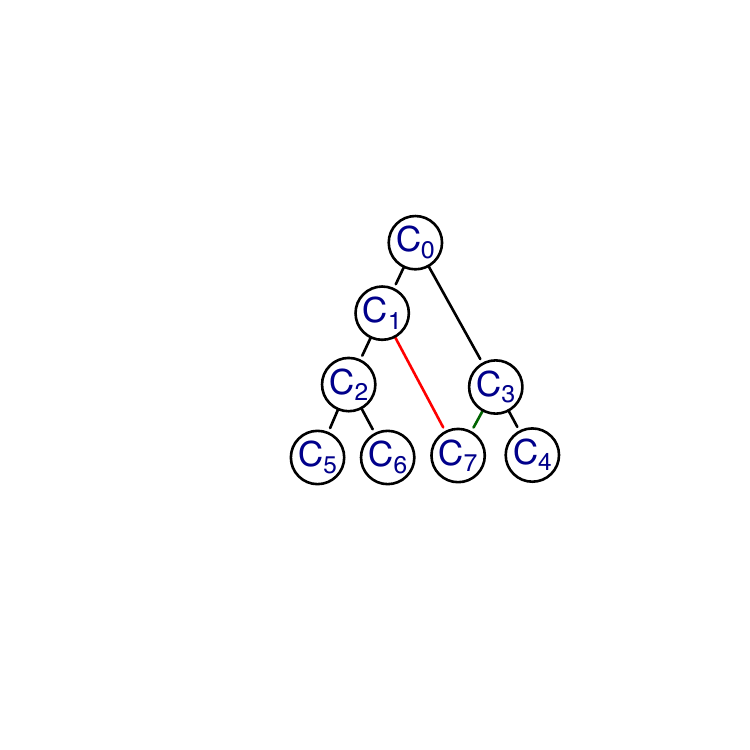} 
\end{center} 
\caption{Polyhierarchy of all link communities of the karate-club network (symbolised by $C_0$). Lower-level communities are subgraphs of higher-level ones. When $C_1$ and the red or $C_3$ and the green link are omitted we obtain one of the two alternative tree-like hierarchies (s.\ text).
} 
\label{Fig-polyhierarchy} 
\end{figure}

In our example, two nested tree-like hierarchies of communities can be observed:\footnote{Hierarchical link clustering also produces tree structures but only some of the many branches of the dendrogram are communities \shortcite{ahn2010link,havemann_identifying_2012}.} 
\begin{enumerate}
\item 
the whole graph splits into $C_1$ and $C_4$ which overlap in boundary node 1, and
 $C_1$ has several communities as subgraphs:
\begin{enumerate}
\item $C_2$ with $C_5$ and $C_6$ as subgraphs,
\item $C_7$,
\end{enumerate}
\item
the whole graph splits into $C_2$ and $C_3$ which overlap in six boundary nodes, and
 each has two communities as subgraphs:
\begin{enumerate}
\item $C_2$ contains  $C_5$ and $C_6$,
\item $C_3$ contains $C_4$ and $C_7$.
\end{enumerate}
\end{enumerate}
$C_3$ is missing in the first hierarchy and $C_1$ in the second. 
These two communities cannot appear in the same nested tree-like hierarchy because they have a permeating overlap, i.e.\ they share not only boundary nodes but also inner nodes. The occurrence of such permeating overlaps is also a new feature of our approach which is absent in the approaches proposed by \shortciteN{ahn2010link} and by \citeN{evans2009line}.
If permeating overlap is a realistic assumption we obtain one solution for the karate-graph but this solution is a polyhierarchy rather than a tree-like hierarchy because $C_7$ is a subgraph of both $C_1$ and~$C_3$ (s.\ Fig.\ \ref{Fig-polyhierarchy}).
In most cases, combined sub-communities do not completely cover the higher-level communities they are part of. The only exception is the whole graph $C_0$, which is completely covered by the pair $(C_1, C_4)$ and also by the pair $(C_2, C_3)$.

\section{Summary and Conclusions}

We propose to see a community being connected via nodes and not via links  to the rest of the network. These boundary nodes can be shared by two or more communities. To evaluate overlapping communities we define a normalised node cut $\Psi$ analogously to usual normalised cut $\Phi$ used to construct disjoint communities. 
We define a community $C$ as a connected subgraph corresponding to a local minimum in a $\Psi$-landscape over all connected subgraphs which are linked by inclusion of a subgraph's neighbour or exclusion of a member. Applying a greedy algorithm, we found seven local $\Psi$-minima of the karate-club network from which nested hierarchies of overlapping communities can be constructed. One pair of communities overlapping in their boundary nodes is compatible with the final split of the karate club. Further tests on benchmark graphs are in preparation. 

Normalised node cut $\Psi$ only uses degrees of nodes as input and can therefore be calculated also for weighted graphs. A greedy algorithm may not find all local minima in the $\Psi$-landscape of connected subgraphs.\footnote{Applying the greedy algorithm to a larger network confirmed this assumption. However, it is possible to identify stable $\Psi$-minima with an evolutionary algorithm. The results will be published elsewhere. (Note added in October 2013)} Whether local $\Psi$-minima represent useful communities depends on the part of reality we model with our network.
\newpage

\section*{Acknowledgements}
We thank Renaud Lambiotte and Steve Gregory for commenting on a draft of this paper. 
This work is part of a project in which we develop methods for measuring the diversity of research. The project is funded by the German Ministry for Education and Research (BMBF). We would like to thank all developers of \textbf{R}.\footnote{\url{http://www.r-project.org}}

\label{appendix}

\appendix
\section{Appendix}
\subsection{Proof of $\Psi(C)=\Phi(L)$}

Normalised node cut $\Psi(C)$ of a subgraph in a network equals normalised edge cut $\Phi(L)$ of the corresponding subgraph in the network's line graph weighted according to \citeN{evans2009line}, as we will show now.

Like in equation \ref{eq:E} we here use $i, j = 1, \ldots, n$ to denote nodes and $k, l = 1, \ldots, m$ for links.
With $L(C)$ we name the set of links between the nodes in $C$. If a link $k$ belongs to $L$ its membership $\mu_k(L)=1$ and zero otherwise.

We calculate the normalised edge cut $\Phi$ of a link set $L$  in the line graph as 
\begin{equation}
\Phi(L)=\frac{K^\mathrm{out}(L)}{K^\mathrm{in}(L)+K^\mathrm{out}(L)}
\end{equation} 
with the sum of internal degrees 
\begin{equation}
\begin{split}
K^\mathrm{in}(L)&=\sum_{k, l = 1}^m \mu_k(L)E_{kl}\mu_l(L)\\
&=\sum_{k, l = 1}^m \mu_k(L)  \sum_{i=1}^n \frac{B_{ik}B_{il}}{k_i}  \mu_l(L)
\end{split}
\end{equation} 
and the sum of external degrees
\begin{equation}
\begin{split}
K^\mathrm{out}(L)&=\sum_{k, l = 1}^m \mu_k(L)E_{kl}(1-\mu_l(L)).\\
&=\sum_{k, l = 1}^m \mu_k(L)\sum_{i=1}^n \frac{B_{ik}B_{il}}{k_i}(1-\mu_l(L)),
\end{split}
\end{equation} 
cf. \shortciteN[eqs. 3 and 4]{havemann_identifying_2012}. 
Now we use the relations
$$\sum_{k=1}^m \mu_k(L) B_{ik} = k_i^\mathrm{in}(C(L))$$
and
$$\sum_{l=1}^m (1-\mu_l(L))B_{il}= k_i^\mathrm{out}(C(L)),$$
which directly follow from the definition of the incidence matrix $B$. Thus, we get 
$$K^\mathrm{in}(L)=\sum_{i=1}^n \frac{(k_i^\mathrm{in}(C))^2}{k_i}$$
and
$$K^\mathrm{out}(L)=\sum_{i=1}^n \frac{k_i^\mathrm{in}(C)k_i^\mathrm{out}(C)}{k_i}.$$
From this we easily derive the sum 
$$K(L) = K^\mathrm{in}(L) + K^\mathrm{out}(L) = \sum_{i=1}^n k_i^\mathrm{in}(C) = k_\mathrm{in}(C) $$
and obtain 
$$ \Phi(L) = \frac{1}{k_\mathrm{in}(C)}\sum_{i \in C}\frac{k_i^\mathrm{in}(C) k_i^\mathrm{out}(C)}{k_i}=\Psi(C),$$
\textit{q.e.d}.

\subsection{Updating $\Psi(C)$}
\label{app.PsiMin}
Let $\sigma(C)$ the sum in the $\Psi$-function and node $i$ a neighbour of $C$. 
For undirected networks the difference $\Delta_i^+\sigma(C) = \sigma(C \cup i) - \sigma(C)$ is given by

$$ \Delta_i^+\sigma(C) =  \sum_{j \in C} A_{ij} \frac{2k_j^\mathrm{out}(C)-A_{ij}}{k_j} - \frac{(k_i^\mathrm{in}(C))^2}{k_i}. $$

The denominator in the $\Psi$-function, $C$'s total internal degree $k^\mathrm{in}(C)$ is increased by $2k_i^\mathrm{in}(C)$
if neighbouring node $i$ is included into $C$. The factor 2 has to be used because in the total internal degree of $C \cup i$  each link is counted two times (for undirected networks). Note, that including neighbour $i$ does not change its internal degree: $k_i^\mathrm{in}(C \cup i) = k_i^\mathrm{in}(C)$ (if there are no self-links). The sum can be restricted to boundary nodes $j \in \beta(C)$ because $A_{ij}=0$ for inner members of $C$. 

If we exclude a boundary node $i$ the numerator $\sigma$ in the $\Psi$-function is changed by  $\Delta_i^-\sigma(C) = \sigma(C \backslash i) - \sigma(C) = - \Delta_i^+\sigma(C \backslash i)$. Because $k_i^\mathrm{in}(C\backslash i)=k_i^\mathrm{in}(C)$ and $k_j^\mathrm{out}(C\backslash i)=k_j^\mathrm{out}(C) + A_{ij}$, we get

$$\Delta_i^-\sigma(C) = \frac{(k_i^\mathrm{in}(C))^2}{k_i} - \sum_{j \in C} A_{ij} \frac{2k_j^\mathrm{out}(C)+A_{ij}}{k_j}.$$

\bibliography{informetrics}

\end{document}